\documentclass[10pt,a4paper,onecolumn]{article}

\usepackage[top=2cm,bottom=2cm,right=4cm,left=4cm]{geometry}
\usepackage{indentfirst}
\usepackage{graphicx} 
\usepackage{float}
\usepackage[utf8]{inputenc}
\usepackage{makeidx}
\usepackage{listings}
\usepackage{mathtools}
\usepackage{pdfpages}
\usepackage{steinmetz}
\usepackage{mleftright}
\usepackage{abstract}
\usepackage{hyperref}

\usepackage{titlesec}
\usepackage[affil-it]{authblk}
\usepackage{blindtext}
\usepackage{todonotes}
\usepackage{subfig}
\usepackage{lineno}
\usepackage{lipsum}

\makeatletter
\newbox\abstract@box
\renewenvironment{abstract}
{\global\setbox\abstract@box=\vbox\bgroup
\hsize=\textwidth\linewidth=\textwidth
\small
\begin{center}%
{\bfseries \abstractname\vspace{-.5em}\vspace{\z@}}%
\end{center}%
\quotation}
{\endquotation\egroup}
\expandafter\def\expandafter\@maketitle\expandafter{\@maketitle
\ifvoid\abstract@box\else\unvbox\abstract@box\if@twocolumn\vskip1.5em\fi\fi}
\makeatother

\titlelabel{\thetitle.\quad}

\renewcommand{\abstractname}{}    
  
\MHInternalSyntaxOn
\renewcommand{\dcases}
{
\MT_start_cases:nnnn
{\quad}
{$\m@th\displaystyle##$\hfil}
{$\m@th\displaystyle##$\hfil}
{\lbrace}
}
\MHInternalSyntaxOff

\begin{document}

\title{\textbf{Shaping High-Order Diffraction-Free Beams Through Continuous Superposition of Bessel Beams}}

\vspace{1 cm}

\author{\normalsize {Michel Zamboni-Rached\thanks{E-mail: mzamboni@decom.fee.unicamp.br}\textsuperscript{1}, Jéssyca Nobre-Pereira\textsuperscript{1} and João Quaglio}\textsuperscript{1}}

\date{\footnotesize{\textsuperscript{1}Universidade Estadual de Campinas, Faculdade de Engenharia Elétrica e Computação, Campinas, SP, Brasil.}}

\begin{abstract}
      
\noindent Recognized for their non-diffracting properties, Bessel beams can be conveniently combined to generate the so-called Frozen Waves, which are monochromatic beams endowed with topological charge and whose longitudinal intensity pattern can be shaped according to a previously chosen function. Continuous superposition of Bessel beams is specially suitable for micrometer-scale domains, being highly relevant for applications in optical tweezers, particle trapping and atom guidance. Previous studies have successfully constructed micrometer Frozen Wave solutions with null topological charge; nevertheless, some challenges persist in obtaining exact solutions when dealing with higher topological charge values (i.e., higher-order micrometer Frozen Waves). Typically, a topological charge raising operator is used to elevate the order; however, such solutions face significant issues when the orders become excessively high. In this paper, based in continuous superposition of higher order Bessel beams, we develop a novel analytic and exact solution for higher-order micrometer Frozen Waves, so obtaining a method for modeling the intensity of beams with arbitrary topological charge within micrometer spatial domains. We also investigate the behavior of the electric field's longitudinal component in such highly non-paraxial regimes through a vectorial treatment. 

      
\end{abstract}

\maketitle

\section{Introduction}

 It is well known that solutions of the cylindrical Helmholtz equation given by $\psi(\rho,\phi,z)=J_\nu(\eta \rho)\exp(i\beta z)\exp(i\nu\phi)$, and known as Bessel beams \cite{Durnin1}, are resistant to diffraction in non-guided media. Here, $\eta=\sqrt{k^2-\beta^2}$ and $\beta$ are the transverse and longitudinal wavenumbers respectively, $k=\omega/c$ is the wavenumber ($\omega$ is the angular frequency and $c$ the light speed), and $\nu$ is the beam's topological charge, also referred as the beam's order. Alongside plane waves, which are also non-diffracting, ideal Bessel beams are not physically realizable since their transverse fields are not square integrable, i.e., they require an infinite energy flux through any transverse plane to maintain their self-healing indefinitely along the propagation. In practice, we can relay on truncated versions \cite{Truncated} or gaussian-apodized versions, as the Bessel-Gauss beams \cite{BG,BG1}. Such type of beams is also referred to as Localized Waves or non-diffracting waves\cite{LocalizedWaves, NonDiffractiveWaves}, and significant progress has been made in this research field, whether in theory, experiments, or applications\cite{localized1,localized2,vortex_bg, Airy}.

Over recent years, superposition of Bessel beams has been used for modelling the longitudinal intensity pattern (LIP) of diffraction-resistant beams. Such method is called \textit{Frozen Wave} (FW), as an allusion to the static envelope \cite{FW1,FW2,experimental,experimental1} of the resulting beams. The Frozen Wave (FW) method has proven versatile, enabling the creation of structured light beams endowed with intriguing properties \cite{StructuredLight, arrays}, even in challenging scenarios like absorbing media \cite{FW3, FW4} and stratified dielectric structures \cite{FW5, FW6}. For instance, numerous studies have explored the theoretical and experimental control of polarization \cite{polarization1, polarization2} and orbital angular momentum density \cite{OAM, OAM1} along the propagation of optical beams. 

The FW beams beams may have applications across various fields, as in optical tweezers \cite{experimental1,NOBREPEREIRA,opt_trap}, atom or particle guidance \cite{guidance}, acoustics \cite{OAM1}, holography \cite{holography} and are also promising in high-resolution microscopy \cite{microscopy}, microfabrication and micromanipulation \cite{microfabri, micromanipulation}.

In this paper, we initially introduce the well-established mathematical method \cite{Y_struct, subluminal_espectro} for deriving a solution for micrometer zero-order FWs through the continuous superposition of zero-order Bessel beams, resulting in a superposition of Mackinnon-type beams \cite{Mackinnon}. However, generating a higher-order FW, i.e., a FW of order $\nu \neq 0$, represents a significant challenge due to the lack of analytical solutions for continuous superposition of higher-order Bessel beams. Given this difficulty, the commonly used approach involves applying an operator, here referred to as the Topological Charge Raising Operator (TCRO), $\nu$ times to the zero-order FW solution to achieve the desired $\nu$-order. Essentially, this method constitutes a recursive differentiation process, and in \cite{NOBREPEREIRA}, we were able to systematize it by presenting the resulting field as a closed mathematical solution in terms of spherical Bessel functions, so enabling the generation of $\nu$ order FWs. However, the TCRO approach proved to work for a limited range of orders, as it modifies the original spectrum (i.e., the spectrum of the original zero-order FW) and thus can affect the resulting Longitudinal Intensity Pattern (LIP).

To address this issue, we introduce a novel method, that we call the Expansion Method, for the generation of FWs of arbitrary order. It employs calculations based on Fourier series expansions, resulting in promising advancements for generating higher-order FWs beams in micrometer spatial regions. Due to the highly non-paraxial behavior of these beams, we also conducted vectorial treatment based on Maxwell's equations to account for the longitudinal component of the electric field. 

The paper is structured as follows: Section 2.1 details the mathematical construction of micrometer zero-order Frozen Waves (FWs). Subsequently, Section 2.2 elucidates the TCRO method, followed by the expansion method and its adaptation for a simple vectorial case in Section 2.3. Both methods are illustrated with examples. Section \ref{conclusions} provides our general remarks and conclusion.


\section{Mathematical Methodology}

\label{mathematical}

\subsection{The FW method based on continuous superposition of zero-order Bessel beams}


An efficient way to obtain optical beams with null topological charge and structured in micrometer regions is through the FW method based on a superposition of forward-propagating and equal frequency Bessel beams with distinct longitudinal (and consequently transverse) wavenumbers $\beta$, that is\footnote{For simplicity, in this paper we will omit the time factor $\exp(-i\omega t)$.} \cite{Y_struct},

\begin{equation}
    \Psi_0(\rho,z)=\int_{-\omega/c}^{\omega/c}S(\beta)J_0(\eta(\beta) \rho)e^{i\beta z}d\beta,
    \label{5}
\end{equation}

\noindent where 

\begin{equation}
    \eta(\beta)=\sqrt{k^2-\beta^2} 
    \label{eta}
\end{equation}

\noindent is the transverse wave number. 

The aim of the method is to obtain, through the superposition \eqref{5}, a beam whose LIP along the $z$ axis is given by the modulus square of a function $F(z)$ of the form 

\begin{equation}
    F(z) = f(z) \mathrm{e}^{iQz} \,\, ,
    \label{F}
\end{equation}

\noindent

where $f(z)$, referred to as the morphological function, is a function of our choice, and the parameter $Q$ is a positive constant, specifically a fraction of $k=\omega/c$, and its role is to ensure that $S(\beta)$ is negligible for $\beta<0$, thereby ensuring that the superposition is primarily composed of forward-traveling waves. Mathematically, we demand that $|\Psi_0(\rho=0,z,t)|^2 \approx |F(z)|^2$, and the FW method demonstrates that this can be achieved by considering the \textit{spectral function} $S(\beta)$ represented by the following Fourier series.

\begin{equation}
    S(\beta) = \sum_{n=-\infty}^{\infty}A_n\exp\left(in\frac{2\pi }{K}\beta \right)\,\,,
    \label{6}
\end{equation}
with 

\begin{equation}
    A_n = \frac{1}{K}F({-2\pi n}/{K})\,\,,
    \label{An}
\end{equation}

\noindent where $K=2\omega/c$. The solution of \ref{5} with the spectrum \ref{6} is given the following superposition of Mackinnon-type beams \cite{Mackinnon}

\begin{equation}\label{eq_order_0}
\Psi_0 \left ( \rho,z,t\right )= \mathrm{e}^{-i \omega t}\sum_{n= - \infty }^{\infty} F\left (-\frac{2n\pi}{K} \right ) \mathrm{sinc}\left [ \xi(\rho,z) \right ]\text{,}
\end{equation}

with: 
\begin{equation}
    \xi(\rho,z) =\sqrt{\frac{\omega^2}{c^2}\rho^2+\left(n\pi+\frac{\omega}{c}z\right)^2}  \text{.}
\end{equation}

Solution \ref{eq_order_0} represents a zero-order Frozen Wave (i.e., $\nu=0$, and thus null topological charge), whose Longitudinal Intensity Pattern (LIP) can be easily modeled in micrometer-scale regions using a function $F(z)$ of our choice. The spot radius of the resulting beam is approximately given by $\rho_{0} \approx 2.4 / \left(k^2 - Q^2\right)^{1/2}$. 


\subsection{The Topological Charge Raising Operator Method}
\label{operatormethod}

In some contexts, such as atom guidance and occasionally in optical tweezers, obtaining solutions for higher-order Frozen Waves ($\nu=1,2,3,...$) becomes essential. Due to the lack of a known method for solving the integral in Eq.(\ref{5}) with (\ref{6}) and the integrand given by a Bessel beam of order $\nu \neq 0$, the procedure used to obtain such higher-order beams involves the utilization of the following operator:

\begin{equation}\label{eq_apend_operator}
U = e^{i\phi}\Big[\frac{\partial}{\partial \rho}+\frac{i}{\rho} \frac{\partial}{\partial \phi}\Big] \text{,}
\end{equation}

\noindent which is referred to here as the Topological Charge Raising Operator (TCRO) due to its ability to increase the order of a beam by one unit if it possesses a well defined topological charge. This method relies on applying the TCRO $\nu$ times to a zero-order Frozen Wave (FW) to obtain a $\nu$-order solution (symbolically, we write $U\Psi_\nu=\Psi_{\nu+1}$). Unlike the case of zero order, the $\nu$-order FW has its intensity pattern distributed over a cylindrical surface with a radius given by 

\begin{equation}\label{rhonu}
\rho_{\nu} \approx \frac{\zeta_{\nu}}{\left(k^2 - Q^2\right)^{1/2}} \text{,}
\end{equation}

 \noindent where $\zeta_{\nu}$ represents the maximum value of $J_{\nu}(\zeta)$ when $\zeta=\zeta_{\nu}$. This is illustrated in Fig. (1b), which shows an example of a second-order FW. 

It has been demonstrated in Ref. \cite{NOBREPEREIRA} that applying the Topological Charge Raising Operator (TCRO) $\nu$ times to the zero-order Frozen Wave, as given by Eq.\ref{eq_order_0}, yields a FW beam of order $\nu$, whose solution is described by the following exact and analytical expression:

\begin{equation} \label{eq_scalar_final}
\begin{split}
     \Psi_{\nu} = (-1)^{\nu} N_{\nu} k^{\nu} e^{i \nu \phi}\sum_{n = -\infty}^{\infty} F\left(-\frac{2n\pi}{K}\right) \Bigg(\frac{\rho}{\sqrt{\rho^2+(z+\frac{n\pi}{k})^2}} \Bigg)^{\nu} j_{\nu}(\xi) \,\,,
\end{split}    
\end{equation}

\noindent where $N_{\nu}=-(k^2-Q^2)^{-\nu/2}/(J_{\nu})_{max}$ is a normalization constant\footnote{The normalization constant $N_{\nu}$ does not emerge naturally in the TCRO application process. We can understand the necessity of this constant as a multiplicative factor in the final solution when we realize that by applying the TCRO $\nu$ times to the integral solution (\ref{5}) of the zero-order beam, two important things occur. The first is the emergence of the factor $(-1)^{\nu}(k^2-\beta^2)^{\nu/2}$ multiplying the original spectrum $S(\beta)$, and the second is the substitution of the zero-order Bessel function, $J_0(\sqrt{k^2 - \beta^2} \rho)$, with the $\nu$-order Bessel function, $J_{\nu}(\sqrt{k^2 - \beta^2} \rho)$ (along with the appearance of the term $\exp(i\nu\phi)$). This causes the resulting beam, although generally having a longitudinal pattern with a shape similar to that of the zero-order beam, to have amplitude values differing by a constant. It can be shown that this constant is approximately given by $N_{\nu}$}, with $(J_{\nu})_{max}$ being the maximum value of $J_{\nu}(\cdot)$, and $j_{\nu}(\cdot)$ is the spherical Bessel function of the first kind.



This technique, of obtaining a higher-order Frozen Wave from a zero-order one through successive applications of the TCRO, has enabled the creation of a broader range of structured micrometer beams endowed with orbital angular momentum, which can be utilized in a variety of applications.

As previously mentioned, in general, the longitudinal intensity pattern of the $\nu$-order Frozen Wave, given by Eq. \eqref{eq_scalar_final}, will exhibit the same Longitudinal Intensity Pattern (LIP) as the original zero-order Frozen Wave, with the intensity shifted from $\rho=0$ to a cylindrical surface of radius $\rho_{\nu}$. However, it is important to highlight that the spectrum of the resulting $\nu$-order beam will not be identical to the spectrum $S(\beta)$ of the original zero-order beam. Instead, it will be given by $(-1)^{\nu}\left(k^2-\beta^2 \right)^{\nu/2} S(\beta)$ (for more details, refer to Ref. \cite{NOBREPEREIRA}) and, consequently, can be considerably different from $S(\beta)$, potentially resulting in undesirable distortions in the Longitudinal Intensity Pattern (LIP) of the higher-order beam. When the original Frozen Wave (zero-order Frozen Wave) has a wide spectrum, a characteristic feature for beams structured in micrometer-scale regions, this distortion effect in the higher-order Frozen Wave given by Eq.\ref{eq_scalar_final} can occur even for small values of $\nu$.

To address the aforementioned limitations associated with the TCRO method, we will introduce in Section 2.3 an exact formulation, referred to here as the expansion method, designed to provide a higher-order Frozen Wave (FW) with the same spectrum as the original zero-order one, making it immune to distortions that may occur when successively applying the TCRO to the initial zero-order beam.

\subsubsection{Example}

In all examples we consider the vacuum and used the following parameters: angular frequency $\omega=3.54\times10^{-15}$ rad/s, implying $\lambda_0 = 0.532 \,\mu \mathrm{m}$ for the wavelength, and $Q=0.85\frac{\omega}{c}$. Considering a morphological function $f(z)$ given by a supergaussian of order 14 centered at $z_{1}=0\,\mu\mathrm{m}$ and width $2*Z_{1}= 20\,\mathrm{\mu m}$, the chosen function $F(z)$ is:

\begin{equation}\label{supergaussiana}
     F(z)=\mathrm{exp}\left (- \left ( \frac{z-z_{1}}{Z_{1}} \right )^{14}  \right )\mathrm{exp}(iQz).
\end{equation}

This example illustrates the application of the TCRO to obtain a second-order micrometer Frozen Wave (FW). Using Eq. \eqref{eq_order_0}, it is possible to construct a zero-order FW with the desired Longitudinal Intensity Pattern (LIP) given by the modulus square of \ref{supergaussiana} and a spot radius of approximately $\rho_0 \approx 0.18 ,\mu\mathrm{m}$, as shown in Fig. (1a). Applying the TCRO twice to this zero-order FW yields the solution \ref{eq_scalar_final} with $\nu=2$, and its intensity is presented in Fig. (1b). It is evident that the LIP is preserved and concentrated over a cylindrical surface with a radius of $\rho_2 \approx 0.49\,\mu\mathrm{m}$.


\begin{figure}[h]
\centerline{\includegraphics[width=15cm]{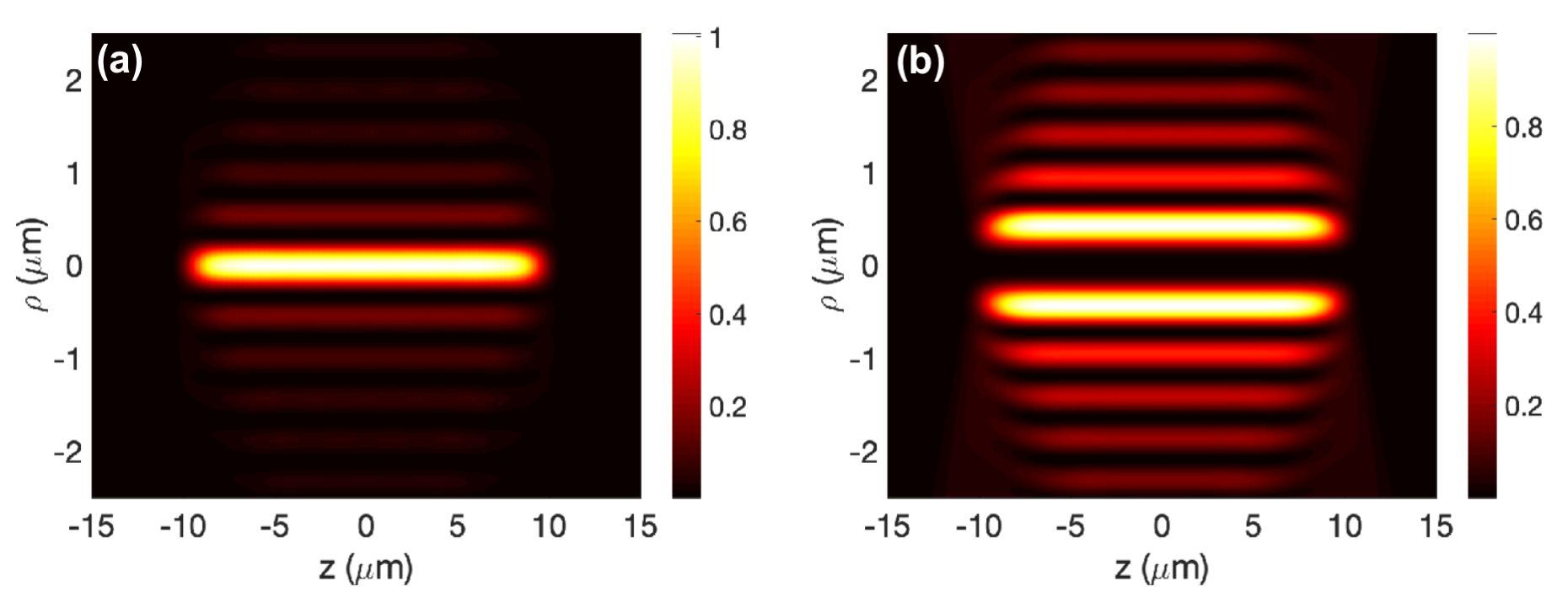}}
\caption{Longitudinal intensity patterns of (a) a zero order FW and (b) second order FW obtained from the TCRO method.}
\label{TCRO_app}
\end{figure}

\subsection{The Expansion Method}
\label{expansionmethod}
\subsubsection{Scalar case}
\label{scalarform}

Our main goal here is to obtain the analytical and exact solution for the following continuous superposition of Bessel beams of order $\nu \neq 0$:

\begin{equation}
\Psi(\rho,\phi,z) = \emph{N}\,e^{i\nu\phi}\int_{-\omega/c}^{+\omega/c}S(\beta)J_{\nu}(\eta(\beta)\rho)e^{i\beta z}d\beta,
\label{Psi2}
\end{equation}

\noindent where $\emph{N}=1/(J_{\nu})_{max}$ and $S(\beta)$ is given by (\ref{6},\ref{An}). With this, we will have a beam of topological charge $\nu$, endowed with the property $|\Psi(\rho=\rho_{\nu},\phi,z)| \approx |F(z)|^2$, where $F(z)$ is given by Eq.(\ref{F}), and $\rho_{\nu}$ by Eq.(\ref{rhonu}). In other words, we will obtain a structured hollow beam with a transverse radius of $\rho_{\nu}$ and a LIP chosen at will, without any risk of obtaining a beam with distortions in the chosen spatial pattern and alterations in the spectrum, as observed in the TCRO method.

Unfortunately, a solution for the integral in (\ref{Psi2}) with $S(\beta)$ given by (\ref{6}) is not known. Now, using an ingenious mathematical trick, we will obtain the desired solution. To do so, let's write the integral solution (\ref{Psi2}) as

\begin{equation}
\Psi(\rho,\phi,z) = \emph{N}\,e^{i\nu\phi}\int_{-\omega/c}^{+\omega/c}\frac{\Lambda(\beta)}{\eta(\beta)}J_{\nu}(\eta(\beta)\rho)e^{i\beta z}d\beta,
\label{Psi_specsolution}
\end{equation}

\noindent where

\begin{equation}
    \Lambda(\beta) = \eta(\beta) S(\beta) \,\, ,
    \label{8}
\end{equation}
\noindent  

\noindent with $\eta(\beta)$ given by Eq.(\ref{eta}).
 
Expanding $\Lambda(\beta)$ into a Fourier series, we have

\begin{equation}
    \Lambda(\beta) = \sum_{n=-\infty}^{\infty}B_n\exp\left(i\frac{2\pi n}{K}\beta\right),
    \label{Lambda}
\end{equation}

\noindent with the coefficients $B_n$ given as

\begin{equation}
    B_n = \frac{1}{K}\int_{-\omega/c}^{+\omega/c}\eta(\beta) S(\beta)e^{-i\frac{2\pi n}{K}\beta}d\beta.
    \label{10}
\end{equation}

Now, by using Eq.(\ref{Lambda}) into Eq.(\ref{Psi_specsolution}) and using the result \cite{grad}

\begin{equation}
\begin{split}
   &\int_{0}^{a}\frac{\cos(cx)}{\sqrt{a^2-x^2}}J_{\nu}(b\sqrt{a^2-x^2})dx \\ &=\frac{\pi}{2}J_{\frac{\nu}{2}}\left[\frac{a}{2}(\sqrt{b^2+c^2}-c)\right]J_{\frac{\nu}{2}}\left[\frac{a}{2}(\sqrt{b^2+c^2}+c)\right],
   \label{expression}
\end{split}
\end{equation}

\noindent we find that the exact analytical solution for Eq.(\ref{Psi2}) is given by

\begin{equation}
\begin{split}
    &\Psi(\rho,\phi,z) = \emph{N}\,\pi e^{i\nu\phi} \\ &\times\sum_{n}B_nJ_{\frac{\nu}{2}}\left[\frac{\omega}{2c}\left(\sqrt{\rho^2+\left(z+\frac{2\pi n}{K}\right)^2}-\left(z+\frac{2\pi n}{K}\right)\right)\right] \\ &\times J_{\frac{\nu}{2}}\left[\frac{\omega}{2c}\left(\sqrt{\rho^2+\left(z+\frac{2\pi n}{K}\right)^2}+\left(z+\frac{2\pi n}{K}\right)\right)\right] \,\,,
    \label{21}
\end{split}
\end{equation}

In order for the solution (\ref{21}) to be fully defined, it is necessary to know the values of the coefficients $B_n$ given Eq.(\ref{10}), whose integral could naturally be solved numerically. However, here we will obtain the analytical solution for it.

We proceed by expanding $S(\beta)$ and $\eta(\beta)$ into Fourier series as follows:

\begin{equation}
S(\beta) = \sum_{m=-\infty}^{\infty}A_m\exp\left(i\frac{2\pi m}{K}\beta\right),
\label{11}
\end{equation}

where, as before,

\begin{equation}
A_m = \frac{1}{K}F\left(\frac{-2\pi m}{K}\right),
\label{12}
\end{equation}

and

\begin{equation}
\eta(\beta) = \sum_{u=-\infty}^{\infty}C_u\exp\left(i\frac{2\pi u}{K}\beta\right),
\label{13}
\end{equation}

with the coefficients $C_u$ determined by

\begin{equation}
C_u = \frac{1}{K}\int_{-\omega/c}^{+\omega/c}\eta(\beta) e^{-i\frac{2\pi u}{K}\beta}d\beta.
\label{14}
\end{equation}

By writing $\eta(\beta)$ as $\frac{\omega}{c}\sqrt{1-x^2}$ where $x^2=\frac{\beta^2}{(\omega^2/c^2)}$, we have that

\begin{equation}
\begin{split}
C_u = \frac{k}{2}\int_{-1}^{1}\sqrt{1-x^2}e^{-iu\pi x}dx \ &= -k\sqrt{\pi}\left(\frac{1}{u\pi}\right)\Gamma\left(\frac{3}{2}\right)J_1(-u\pi)\,\,,
\end{split}
\label{C_u}
\end{equation}

\noindent where $\Gamma(\cdot)$ is the gamma function.

Finally, by using Eq. (\ref{11}) into Eq. (\ref{10}), we have

\begin{equation}
B_n = \sum_{m=-\infty}^{\infty}A_m\frac{1}{K}\int_{-\omega/c}^{+\omega/c}\eta(\beta) e^{-i\frac{2\pi(n-m)}{K}\beta}d\beta,
\end{equation}

and so

\begin{equation}\label{Bncomum}
B_n = \sum_{m}A_mC_{n-m}.
\end{equation}

With this, using Eqs.(\ref{12},\ref{C_u}), the coefficients $B_n$ are given by

\begin{equation}\label{Bn}
B_n = \sum_{m=-\infty}^{\infty} \, -\frac{1}{2\sqrt{\pi}(n-m)} \Gamma\left(\frac{3}{2}\right) F\left(\frac{-2\pi m}{K}\right) J_1(-(n-m)\pi)\,.
\end{equation}

To summarize, a Frozen Wave of order $\nu \neq 0$, obtained by the continuous superposition of Bessel beams through Eq.(\ref{Psi2}) with $S(\beta)$ given by Eqs. (\ref{6},\ref{An}), is described by the exact analytical solution provided in Eq. (\ref{21}), where the coefficients $B_n$ are determined by Eq.(\ref{Bn}).

\subsubsection{Vectorial case}

Micrometer optical Frozen Waves are shaped in such small dimensions (nonparaxial regime) that a scalar analysis of the phenomenon is not appropriate. Taking this into account, we will adopt a vectorial approach here, considering an optical beam whose electric field is given by

\begin{equation}\label{vec_form}
\textbf{E}=E_{0}\Psi(\rho,\phi,z)\mathbf{\hat{x}}+E_{z}\mathbf{\hat{z}},
\end{equation}

\noindent Here, the spatial modeling is performed on the transverse component of the field, $E_x=E_{0}\Psi(\rho,\phi,z)$, with $\Psi(\rho,\phi,z)$ given by the solution (\ref{21}) from Section 2.3.1, and $E_0$ is a constant with a value of one and units of electric field. The axial component $E_z$ is then obtained from Gauss's law:

\begin{equation}\label{Ez_int}
 E_z=-E_{0}\frac{\partial }{\partial x}\int \Psi(\rho,\phi,z)dz,
\end{equation}

\noindent where $\frac{\partial}{\partial x}$ will be written later in cylindrical coordinates.
After inserting Eq. \eqref{21} into Eq. \eqref{Ez_int}, it is evident that the integral with respect to $z$ is challenging to perform analytically. To address this issue, we use a strategy involving the spectral representation of  $\Psi(\rho,\phi,z)$ provided by Eq. \eqref{Psi_specsolution}. This approach simplifies the integration, as the only term depending on $z$ in this expression is $\mathrm{e}^{i\beta z}$. Consequently, after using (\ref{Psi_specsolution}) into \eqref{Ez_int}, we get 

\begin{equation}\label{Ez_int1}
E_z=-\emph{N}\,E_{0}\frac{\partial }{\partial x} \left [   e^{i\nu\phi}\int_{-\omega/c}^{+\omega/c}\frac{\Lambda(\beta)}{i \beta\eta(\beta)}J_{\nu}(\eta(\beta)\rho)e^{i\beta z}d\beta \right ]
\end{equation}

\noindent with $\Lambda(\beta)$ given by Eq. \eqref{8} and $\eta(\beta)$ by Eq.(\ref{eta}). 

To start with the evaluation of the integral appearing in Eq.(\ref{Ez_int1}), let us express $\eta(\beta)S(\beta)/i\beta$ in Fourier series

\begin{equation}
    \eta(\beta) \frac{S(\beta)}{i\beta} = \sum_{n=-\infty}^{\infty}B'_n\exp\left(i\frac{2\pi n}{K}\beta\right)\,\, ,
    \label{Lambda_exp}
\end{equation}

\noindent with the coefficients $B'_n$ given by

\begin{equation}
    B_n' = \frac{1}{K}\int_{-\omega/c}^{+\omega/c}\eta(\beta) \frac{S(\beta)}{i\beta}e^{-i\frac{2\pi n}{K}\beta}d\beta.
    \label{Bn_comma}
\end{equation}

By using Eq.(\ref{Lambda_exp}) into Eq.(\ref{Ez_int1}), and using the result given by Eq.(\ref{expression}), we obtain 

\begin{equation}\label{Ez_comp}
\begin{split}
&E_z(\rho,\phi,z) = \left( \cos\phi\frac{\partial}{\partial\rho}-\frac{\sin\phi}{\rho}\frac{\partial}{\partial\phi} \right) \left\{ N\,\pi \mathrm{e}^{i\nu\phi}\times \right.\\
&\left. \sum_{n}B_n'J_{\frac{\nu}{2}}\left[\frac{\omega}{2c}\left(\sqrt{\rho^2+\left(z+\frac{2\pi n}{K}\right)^2}-\left(z+\frac{2\pi n}{K}\right)\right)\right] \right.\\ 
&\left. \times J_{\frac{\nu}{2}}\left[\frac{\omega}{2c}\left(\sqrt{\rho^2+\left(z+\frac{2\pi n}{K}\right)^2}+\left(z+\frac{2\pi n}{K}\right)\right)\right] \right\}\,\,,
\end{split}
\end{equation}


\noindent where we have written $\frac{\partial}{\partial x}$ in cylindrical coordinates.

Equation (\ref{Ez_comp}) is the exact analytical solution for the axial component of the electric field of the FW beam, but for it to be complete, it is necessary to know the values of the coefficients $B'_n$ given by Eq.(\ref{Bn_comma}), whose integral has to be solved. To do this, we will use the that

\begin{equation}\label{spec_comma}
\frac{S(\beta)}{i\beta} =  \sum_{m=-\infty}^{\infty}A_m'\exp\left(i\frac{2\pi m}{K}\beta\right),
\end{equation}

\noindent with

\begin{equation}
    A_m' = \frac{1}{K}g\left(\frac{-2\pi m}{K}\right),
    \label{Am_linha}
\end{equation}

\noindent and

\begin{equation}\label{g_coeff}
g\left ( \frac{-2\pi m}{K} \right )=\left [ \int F(z)dz \right ]_{z=-\frac{2n\pi}{K}}.
\end{equation}

Now, by using Eq. \eqref{spec_comma} into Eq. \eqref{Bn_comma} we have

\begin{equation}
    B_n' = \sum_{m=-\infty}^{\infty}A_m'\frac{1}{K}\int_{-\omega/c}^{+\omega/c}\eta(\beta) e^{-i\frac{2\pi(n-m)}{K}\beta}d\beta,
\end{equation}

\noindent then

\begin{equation}\label{Bncomum_comma}
    B_n' = \sum_{m}A_m'C_{n-m}.
\end{equation}

\noindent where $C_{n-m}$ can be obtained from Eq. \eqref{C_u}. Therefore, the final expression of $B_n'$ is given by

\begin{equation}\label{Bnlinha}
    B_n' = \sum_{m} \, -\frac{1}{2\sqrt{\pi}(n-m)} \Gamma\left(\frac{3}{2}\right) g\left(\frac{-2\pi m}{K}\right) J_1(-(n-m)\pi)\,\, ,
\end{equation}

\noindent with $g\left(\frac{-2\pi m}{K}\right)$ given by Eq.(\ref{g_coeff}).

\subsubsection{Example}

\textit{Example 1.} Here, we construct two scalar sixth-order FWs with the LIP dictated by the function given by Eq.(\ref{supergaussiana}), where the first one, Fig. (2a), is obtained using the TCRO method, and the second one, Fig.(2b), using the expansion method. In both cases, we use the same wavelength and parameter $Q$ values as in the example from Section 2.3.1.

It is evident that the FW obtained with the TCRO method exhibits distortions at the edges. This occurs because, as previously mentioned, the TCRO method can reproduce the intended LIP within certain limitations. In this case, the six successive applications (since $\nu=6$) of the TCRO on the zero-order FW are sufficient to considerably alter the spectrum of the resulting beam, leading to the distortion in the desired LIP.

On the other hand, we observe that the FW obtained by the expansion method is immune to such distortions, providing the desired LIP. This is due to the fact that the spectrum of the resulting sixth-order beam is, in fact, the spectrum $S(\beta)$.



\begin{figure}[h]
\centerline{\includegraphics[width=15cm]{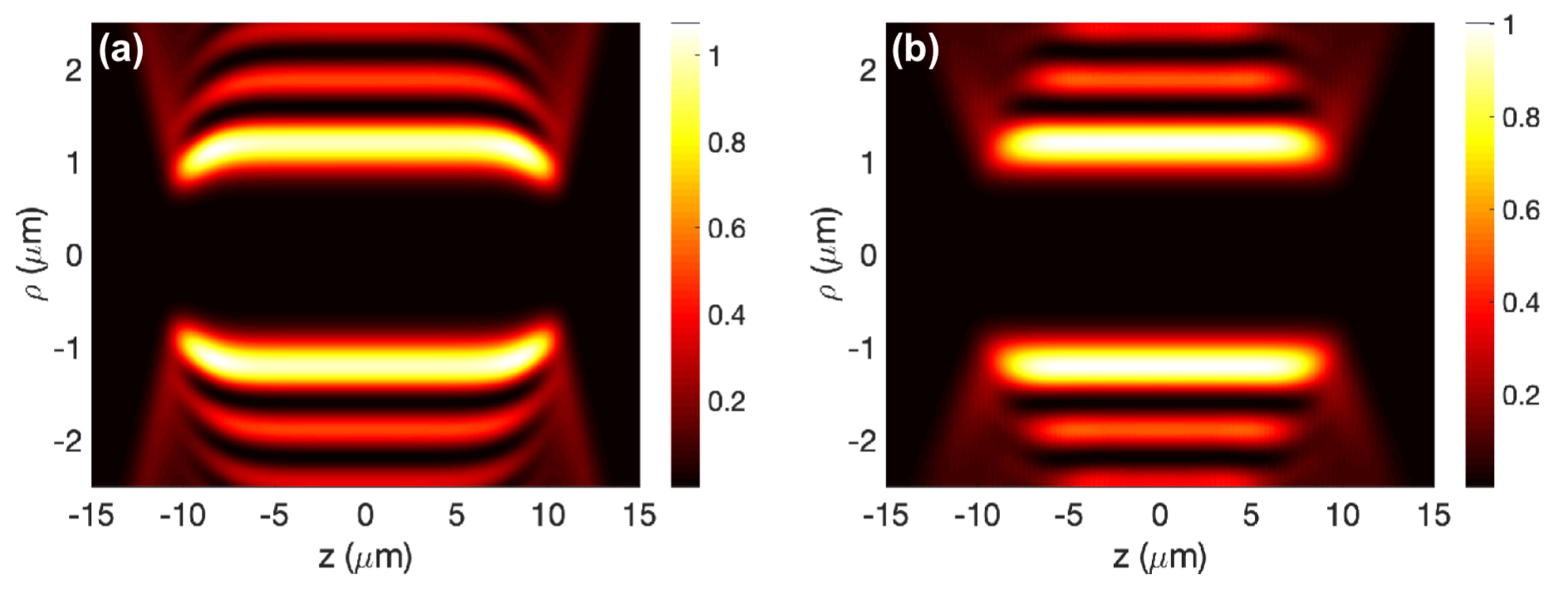}}
\caption{Intensities of two 6th-order scalar FWs designed to have the same LIP, with: (a) FW obtined from the TCRO method; (b) FW obtained from the expansion method.}
\label{comparison_methods}
\end{figure}

\textit{Example 2.} In this example, we apply the expansion method considering the vectorial nature of the beam, more specifically considering a beam with an electric field given by Eq. (\ref{vec_form}), with $\Psi$ and $E_z$ given by Eqs. (\ref{21}) and (\ref{Ez_comp}), respectively, where the transverse component $E_x=\Psi$ is the same sixth-order FW from the previous example obtained by the expansion method.
Figures (3a) and (3b) show, respectively, the 3D intensities of the transverse and axial components of the electric field, with the inset figures displaying the intensity fields on the $z=0$ plane. As mentioned earlier, for highly non-paraxial beams, a longitudinal component emerges, as observed in Figure (3b) and, in some cases, it can present a non-negligible part in the composition of the resulting beam.

\newpage


\begin{figure}[h]
\centerline{\includegraphics[width=15cm]{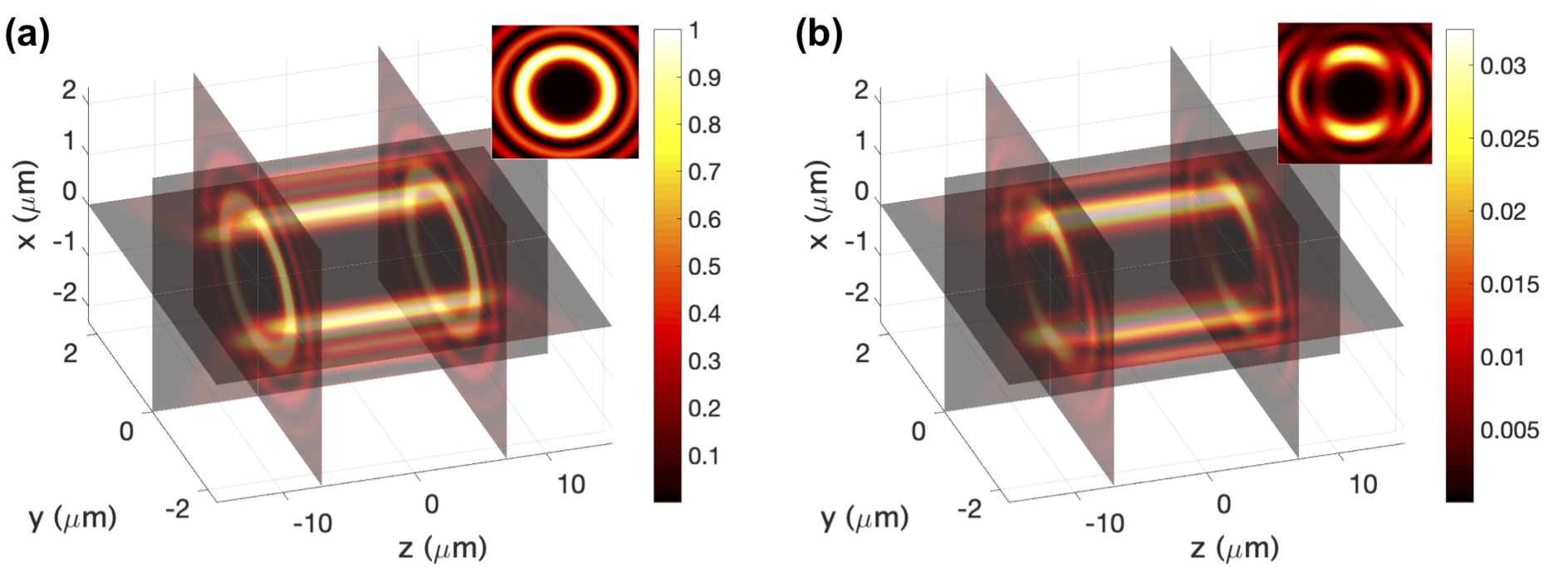}}
\caption{a) 3D intensity of the transverse electric field of the FW beam obtained from the expansion method; b)  3D intensity of the axial component $E_z$ of the electric field of the FW beam obtained from the expansion method. Insets: intensity fields on the $z=0$ plane.}
\label{ExEzcomponent_Exact}
\end{figure}

\section{Conclusions}
\label{conclusions}

The continuous superposition of Bessel beams is a powerful tool for spatial modeling of the optical field in micrometer regions. This methodology, known as the Frozen Wave method, is well-developed for the zero-order case, i.e., the superposition of zero-order Bessel beams, providing analytical and exact solutions for structured light beams with zero topological charge. However, applying the method for higher orders, aiming to obtain analytical and exact solutions that provide structured beams with orbital angular momentum, is mathematically more complex. The only known method to mitigate such difficulties was the use of a Topological Charge Raising Operator (TCRO). By applying this operator $\nu$ times to a structured zero-order beam, it can produce a structured beam of order (topological charge) $\nu$. However, the successive application of the TCRO affects the spectrum of the resulting beam, leading to distortion in the desired spatial pattern.

Addressing this issue, we have developed what we refer to as the "expansion method," which definitively resolves the challenges of the Frozen Wave method for higher orders. It provides an analytical and exact approach to obtaining structured beams of arbitrary order (topological charge) in micrometer regions. Furthermore, considering that structured beams in micrometer regions are highly non-paraxial, we extended the expansion method to the vectorial case, taking into account the axial component of the electric field. This method is particularly relevant for research in atom guidance, optical trapping, micro-manipulation of particles, microlithography, imaging, and medical applications.

\end{document}